\documentclass[twocolumn,prd,showpacs,nofootinbib,a4]{revtex4-1}
\usepackage{graphicx}
\usepackage{enumerate}
\usepackage{amsmath}
\usepackage{hyperref}
\usepackage{breakurl}

\def\ppbar{p/$\bar{\rm p}$}
\def\pbar{$\bar{\rm p}$}

\begin{document}

\title{CMB bounds on dark matter annihilation: Nucleon energy-losses after
recombination}

\author{Christoph Weniger$\,{}^{a}$, Pasquale D.~Serpico$\,{}^{b}$, Fabio
Iocco$\,{}^{c}$, Gianfranco Bertone$\,{}^{a}$}
\affiliation{$^{a}$GRAPPA Institute, Univ.~of Amsterdam, Science Park 904,
1098 GL Amsterdam, Netherlands}
\affiliation{$^{b}$LAPTh, Univ.~de Savoie, CNRS, B.P.110, Annecy-le-Vieux
F-74941, France}
\affiliation{$^{c}$The Oskar Klein Center for CosmoParticle Physics,
Department of Physics, Stockholm University, Albanova, SE-10691 Stockholm,
Sweden}

\date{\today}

\begin{abstract}
  We consider the propagation and energy losses of protons and anti-protons
  produced by dark matter annihilation at redshifts $100<z\lesssim2000$.  In
  the case of dark matter annihilations into quarks, gluons and weak gauge
  bosons, protons and anti-protons carry about 20\% of the energy injected
  into $\rm e^{\pm}$ and $\gamma$'s, but their interactions are normally
  neglected when deriving cosmic microwave background bounds from altered recombination histories.
  Here, we follow numerically  the energy-loss history of typical
  protons/antiprotons in the cosmological medium. We show that about half of
  their energy is channeled into photons and $\rm e^{\pm}$, and we present  a
  simple prescription to estimate the corresponding strengthening of the
  cosmic microwave background
  bounds on the dark matter annihilation cross section.
\end{abstract}
\pacs{95.30.Cq, 
95.35.+d, 
98.80.Es
\hfill  LAPTH-007/13}

\maketitle

\section{Introduction}

Astrophysical and cosmological observations provide compelling evidence that
about 85\% of all the matter in the Universe is in the form of Dark Matter
(DM), an elusive substance which is currently searched for with a variety of
observational and experimental channels at colliders, in underground
detectors, or via indirect signals from DM annihilation or decay
\cite{reviews}.

Cosmic microwave background (CMB) anisotropy and polarization data provide
interesting constraints on the properties of DM
particles~\cite{Chen:2003gz,Padmanabhan:2005es,Kanzaki:2008qb}. Secondary
particles injected via DM annihilation (or decay) after recombination, around
redshift $z\sim {\cal O}(600)$, would in fact inevitably affect the
recombination history of the Universe and widen the surface of last
scattering, which is tightly constrained by CMB observations as discussed in
Refs.~\cite{Galli:2009zc,Slatyer:2009yq,Huetsi:2009ex,Cirelli:2009bb,
Kanzaki:2009hf}, and more recently in
Refs.~\cite{Hutsi:2011vx, Galli:2011rz,Finkbeiner:2011dx,Giesen:2012rp,Evoli:2012qh,
Slatyer:2012yq, Cline:2013fm}. Possible effects on the cosmological
recombination spectrum were discussed in Ref.~\cite{Chluba:2009uv}.

One of the main reasons of interest for these constraints is that, in contrast
with other indirect searches, they do not rely on knowledge of astrophysical
DM structures, affected by the complex aspects of non-linear gravity as well
as complicated feedback due to baryons~\cite{Huetsi:2009ex, Cirelli:2009bb}. The CMB probe is thus as reliable as
the description of the basic atomic and nuclear/particle physics processes
involved is.  The robustness (and astrophysical independence) of the CMB constraints motivates further
efforts to assess and improve the error budget.
The degree of sophistication in modeling the atomic processes down to the recombination
stage is quite elevated and has also seen recent improvements, see e.g.~\cite{Chluba:2010ca}.
Here we revise one aspect related to the accuracy of the nuclear/particle physics part.

It is typically assumed that protons are highly penetrating and poor at
transferring energy to the intergalactic medium (IGM) --- a misnomer, since no
galaxies have formed at such high redshifts ---  and their energy release to
the medium is neglected~\cite{Chen:2003gz,Slatyer:2009yq} (see however
the comment in Ref.~\cite{Hutsi:2011vx}). In this article,
we estimate the additional energy released to the gas by the interactions of
the high-energy protons and antiprotons formed among DM annihilation final
states.  

This article is structured as follows: In Sec.~\ref{processes} we review the
basic physics substantiating the two points above. In Sec.~\ref{computation}
we describe our computational technique and present our results in
Sec.~\ref{results}. Finally, in Sec.~\ref{conclusions} we conclude.

\section{Physical processes}\label{processes}
Protons and antiprotons carry a significant fraction of the overall energy
emitted in DM annihilations into quarks, gluons or weak gauge bosons.
Typically, this amounts to $\sim20$\% of the energy channelled into $\rm
e^{\pm}$ and $\gamma$'s, see for example Fig.~4 in~\cite{Cirelli:2010xx}, and
a fraction of this energy will be inevitably transferred to the IGM.  Neutrons
and antineutrons decay very fast and behave practically like protons and
antiprotons, while (anti-)deuterons and heavier nuclei are produced in
negligible amounts in DM annihilations.

At the epochs of interest here, which correspond to a redshift
$z=\mathcal{O}$(10$^3$), \ppbar's propagate in a medium which is  eight or
nine orders of magnitude more dense than at the present epoch, with typical proton
and Helium gas densities up to $\mathcal{O}$(10$^2$) particles/cm$^3$. Even
neglecting the interaction with photon baths of densities of ${\cal
O}(10^{11})$cm$^{-3}$, a typical p-p inelastic cross-section of 30 mb yields a
collision timescale lower than the age of the Universe at decoupling. Hence,
we expect that in general the probability for a nucleon to interact  within a
Hubble time is large, and that a significant energy deposition takes place
(similar estimates can be found in Ref.~\cite{Hutsi:2011vx}).

More specifically, \ppbar's will undergo the following processes:
\begin{itemize}
  \item Coulomb scattering of \ppbar~on IGM
    electrons~\cite{Jackson:1998}\footnote{As lower cutoff in the Coulomb
    logarithm we adopt $10\ \rm eV$.}
  \item Thomson scattering of \ppbar~on IGM photons~\cite{Jedamzik:2006xz}
  \item Inelastic scattering of \ppbar~on IGM protons~\cite{Moskalenko:2001ya}
  \item Inelastic scattering of \ppbar~on IGM
    Helium~\cite{Moskalenko:2001ya}.
\end{itemize}

\begin{figure}[t]
  \begin{center}
    \includegraphics[width=\columnwidth]{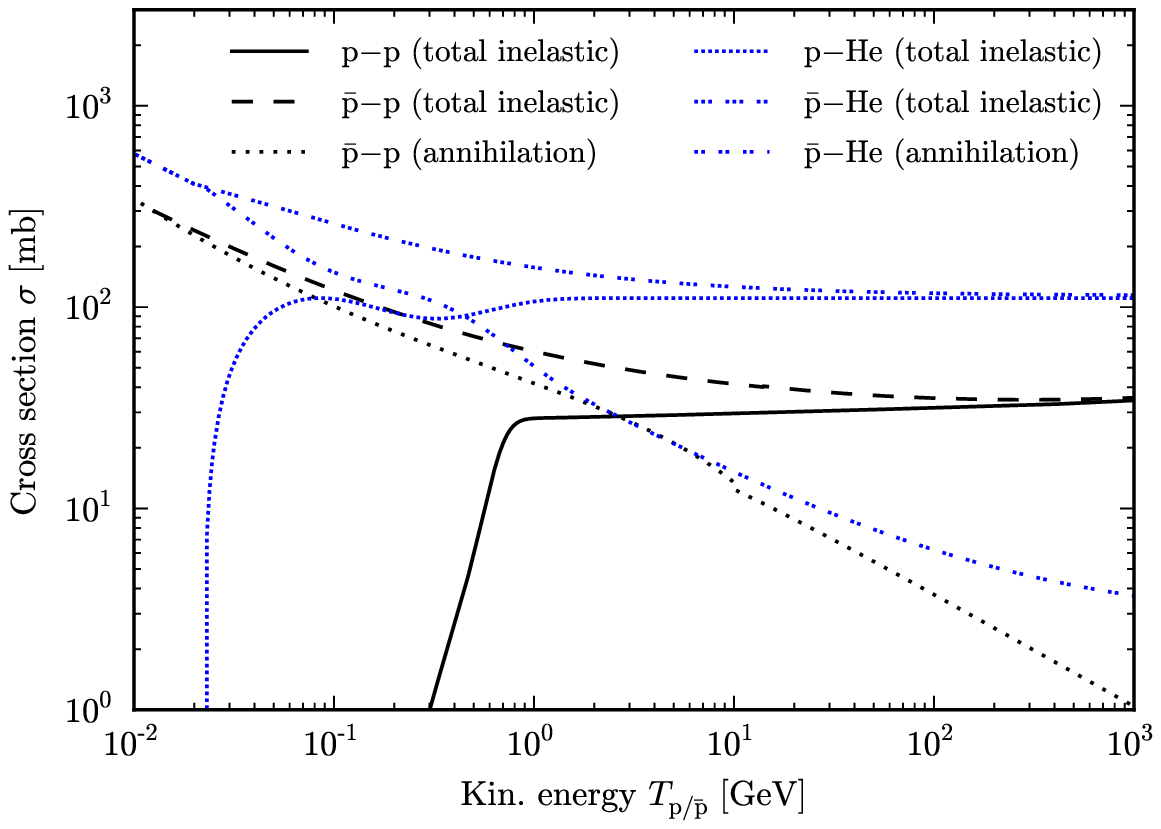}
    \includegraphics[width=\columnwidth]{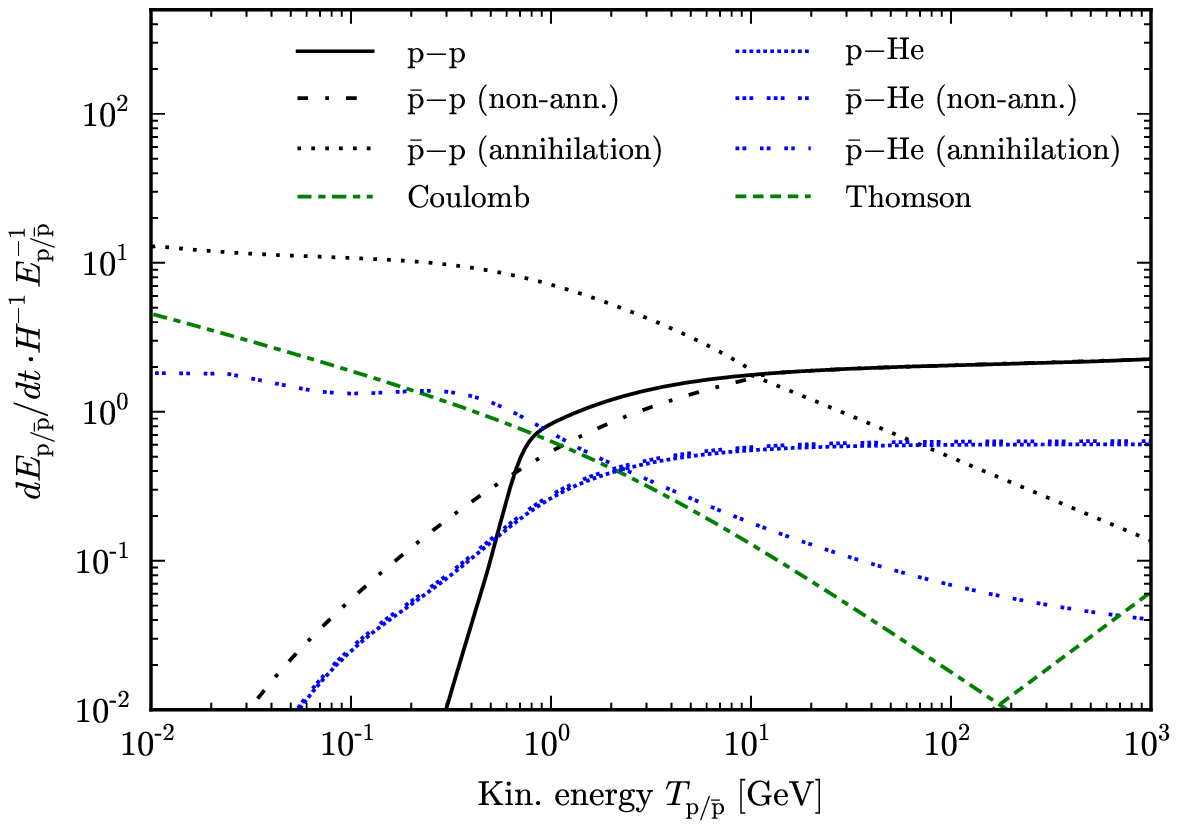}
  \end{center}
  \caption{\emph{Upper panel:} Total inelastic hadronic cross-sections in lab
  frame as function of the $p$ and $\bar p$ kinetic energies.  Where
  applicable, we also show the annihilation part of the total cross-section
  separately. \emph{Lower panel:} Fractional energy losses relative to Hubble
  rate, at redshift $z=1000$. The non-annihilating and annihilating parts of
  the hadronic cross-sections are shown separately.}
  \label{fig:losses}
\end{figure}

In Fig.~\ref{fig:losses} we show the inelastic scattering cross-sections as a
function of the \ppbar~kinetic energy (top panel), and the corresponding
fractional energy loss rate $E^{-1}dE/dt$ times the Hubble time $H^{-1}$ at
redshift $z=1000$ as function of the \ppbar~kinetic energy (bottom panel).
Values of order unity or even larger indicate potentially large effects. In
case of anti-protons, we plot annihilating and non-annihilating rates
separately. For \emph{non-annihilating} inelastic interactions, the final
energy distribution of \ppbar~is taken to be constant in the physically
accessible regime $m_p$\dots$E_p$ ($E_p$ is the energy of the primary
particle, $m_p$ the \ppbar~mass). Consequently, the mean energy loss during a
non-annihilating collision is $\langle\Delta E\rangle \simeq0.5 T_p$, with
$T_p\equiv E_p-m_p$ being the initial kinetic energy. The energy released
during an \emph{annihilation} event is $\Delta E = E_p+m_p$.  At energies
$E\gtrsim10\rm\ GeV$, the losses are dominated by inelastic \pbar-p and p-p
scattering without annihilation; at lower energies \pbar-p annihilation and
Coulomb losses become relevant and dominate for non-relativistic particles.

In the next section we shall follow the evolution of a nucleon pair from a
hypothetical  toy annihilation process $\chi\chi\to \rm \bar p p$, computing
the energy loss of the daughter nucleons down to relatively low-redshifts of a
few hundreds.  Obviously, only part of this energy will be {\it absorbed} by
the gas.  Following in detail the energy degradation down to the
energy-transfer processes to the gas --- heating and ionization --- goes
beyond our purposes (it is worth mentioning that several efforts are being put
into a more realistic treatment of the related physics for the $\rm e^{\pm}$
and $\gamma$'s, see e.g.~\cite{Evoli:2012zz}).  Rather, we will content
ourselves with estimating the energy that is released into energetic electrons
and photons (essentially, the fraction of the energy lost to stable particles
other than neutrinos). We will express this as the \textit{electromagnetic
fraction} $f_{e^\pm\gamma}$ of the energy initially injected as nucleons,
which we will precisely define below. This approach is exact as long as the
CMB bounds are ``calorimetric'' and the precise form of the spectrum of
electromagnetic particles is irrelevant in determining the ultimate fate of
the energy injected. 

Previous investigations suggest that this should provide a reasonable first
approximation to the true result; for our purposes the accuracy is expected to
be at the level of $\sim30\%$ (see Fig.~2 of \cite{Slatyer:2012yq},
$z\sim600$).  Since we are already concerned with a correction to the basic
results, we shall adopt this Ansatz which greatly simplifies the problem.  The
only potentially problematic case is the one of Coulomb reactions of the
(anti-)protons, which can produce low-energy electrons (which may behave
differently from high-energy ones). However, the legitimacy of our
approximation is supported by two arguments: i) this energy-loss rate has only
a logarithmic dependence on the minimum kinetic energy; ii) it only matters
for non-relativistic (anti-)protons, for which most of the energy coming from
electroweak-scale WIMPs has already been dissipated. In fact, with the
exception of very light (GeV scale) DM particles, protons are typically born
relativistic or semi-relativistic.

\section{Computation}\label{computation}
The energy loss processes sketched above are of two distinct classes:
continuous energy losses (with small energy transfer per collision) when
protons interact with electrons or photons; or catastrophic losses, when they
undergo an inelastic collision or an annihilation (for antiprotons). For the
moment we shall ignore elastic collisions, we will comment on their role at
the end of Sec.~\ref{results}.

Continuous energy losses in the range of our interest, $100<z\lesssim2000$,
can be described via the following differential equation
\begin{equation}
  \frac{dE}{dz}=\left.\frac{dE}{dz}\right|_{\rm
  Coulomb}+\left.\frac{dE}{dz}\right|_{\rm Thomson}+ \frac{v^2E}{1+z}\;.
\end{equation}
The last term at the RHS describes adiabatic energy-losses in the expanding
universe, the other two terms describe Coulomb and Thomson energy-losses. In
the present work, Thomson losses will be neglected, which from
Fig.~\ref{fig:losses} is clearly justified at the energies of interest.

Hadronic processes are included on top of the continuous energy losses with a
Monte Carlo simulation. The Monte Carlo starts with a fixed \ppbar~energy $E_{\rm
inj}$ at some initial redshift $z_{\rm init}$ and tracks the history of the
particle as it moves to lower redshifts.  The redshift of an individual
scattering process is inferred by sampling from a survival function, which is
determined by solving a differential equation whose derivate is given by the
inelastic scattering rates. The amount of electromagnetic and hadronic energy
lost by each particle as function of redshift $z$ is recorded. 

At any given redshift $z$, a fraction $f_{e^\pm\gamma}(z)$ of the hadronically
injected energy is actually channeled into electromagnetic form (energetic
$e^\pm$ and photons, see discussion above).  More specifically, for
annihilating DM, $f_{e^\pm\gamma}$ is defined as
\begin{equation}
  \epsilon_{e^\pm\gamma}(z)\equiv\frac{dE}{dVdt} = 2 M_\chi
  f_{e^\pm\gamma}(z)\cdot \langle\sigma v\rangle n^2_{\chi,0}\cdot (1+z)^3\;,
  \label{}
\end{equation}
where $\epsilon_{e^\pm\gamma}(z)$ denotes here the energy released into
electromagnetic form \emph{per comoving volume} and per unit time, and
$n_{\chi,0}$ is today's number density of DM particles.  This
$f_{e^\pm\gamma}(z)$ can be derived from the integral
\begin{equation}
  f_{e^\pm\gamma}(z) = 
  \int_z^{z_{\rm max}} dz'\ g_{e^\pm\gamma}(z,z')\
  \frac{(1+z')^2}{H(z')(1+z)^3}\;,
  \label{eqn:g2f}
\end{equation}
where $g_{e^\pm\gamma}(z,z')$ denotes the fraction of the energy of a single
DM annihilation event at redshift $z'$ that is released into electromagnetic
energy at redshift $z$ per unit time.  We will take $z_{\rm max}=10000$
throughout, such that $\epsilon_{e^\pm\gamma}(z)$ at $z<1000$ is independent
of $z_{\rm max}$. 

We split $f_{e^\pm\gamma}(z)$, and analogously the function
$g_{e^\pm\gamma}(z)$, into three parts according to
\begin{equation}
  f_{e^\pm\gamma} = \tilde f_{\rm Cou.} + \alpha_{\rm had} (\tilde f_{\rm
  inel.} + \tilde f_{\rm ann.})\;,
  \label{eqn:split}
\end{equation}
with the three addenda at the RHS corresponding to Coulomb losses, inelastic
scattering and annihilation, respectively. These $\tilde f$'s denote the total
(anti-)proton energy \emph{lost} to secondary particles via the scatterings.
All of the energy lost by Coulomb scattering, and a fraction $\alpha_{\rm
had}$ of the energy lost during inelastic scattering and annihilation is
channeled into electromagnetic energy and hence contributes to
$f_{e^\pm\gamma}$.
\smallskip

A reasonable value is $\alpha_{\rm had}\simeq0.5$.  After a hadronic
scattering, the energy eventually carried by $e^{\pm}$ and $\gamma$
contributes fully to $f_{e^\pm\gamma}$, the energy carried by $\nu$'s is
completely lost, and only a fraction of the energy carried by nucleons will be
transfered into $e^\pm\gamma$ at a later stage.  We will neglect this latter
contribution to $f_{e^\pm\gamma}$, since it is further suppressed by the small
fraction of energy released into nucleons during inelastic scattering.

As a proxy for the energy released in the different channels we can take the
gluon-gluon (or charmed quark) channels of Fig.~4 in~\cite{Cirelli:2010xx},
where about 46\% of the energy is found to be channelled into $e^{\pm}$ and
$\gamma$, with $\sim 27\%$ of this amount going into nucleons. We also checked
that these figures do not change appreciably as a function of $M_\chi$ (M.
Cirelli, private communication). This is also consistent with the expectations
from non-relativistic proton-antiproton annihilation, see e.g.~Table 2
in~\cite{Steigman:1976ev}.  A very simple justification of this value can be
obtained in the limit where annihilation final states are just made of pions,
with isospin-blind production yields (equal quantities of
$\pi^0,\pi^+,\pi^-$). One would expect full channeling of $\pi^0$ energy into
$e^\pm\gamma$ and about $1/6$ of the energy stored in $\pi^\pm$, hence a
weighted average of
44\%. We thus believe that $\alpha_{\rm had}\simeq 0.5$ is a reasonable
benchmark, probably affected by a $20\%$ uncertainty. While this may be a
crude description of the actual process, it is enough for providing a first
estimate of the hadronic correction to CMB bounds on annihilating dark matter.

On a more technical note, to obtain $\tilde g_{\rm Cou.}(z,z')$, $\tilde
g_{\rm inel.}(z,z')$ and $\tilde g_{\rm ann.}(z,z')$, we simulate particle
injection at redshifts $z'=30$--$10000$ and record what fraction of the energy
is released in certain redshift intervals $z_0\dots z_1$, divided by the
corresponding time interval $\Delta t\equiv t_0 - t_1$.  From these $\tilde
g(z, z')$'s, we can then derive the $f_{e^\pm\gamma}(z)$ from
Eqs.~\eqref{eqn:g2f} and~\eqref{eqn:split}.

\section{Results and Discussion}
\label{results}
\begin{figure}[t]
  \begin{center}
    \includegraphics[width=\columnwidth]{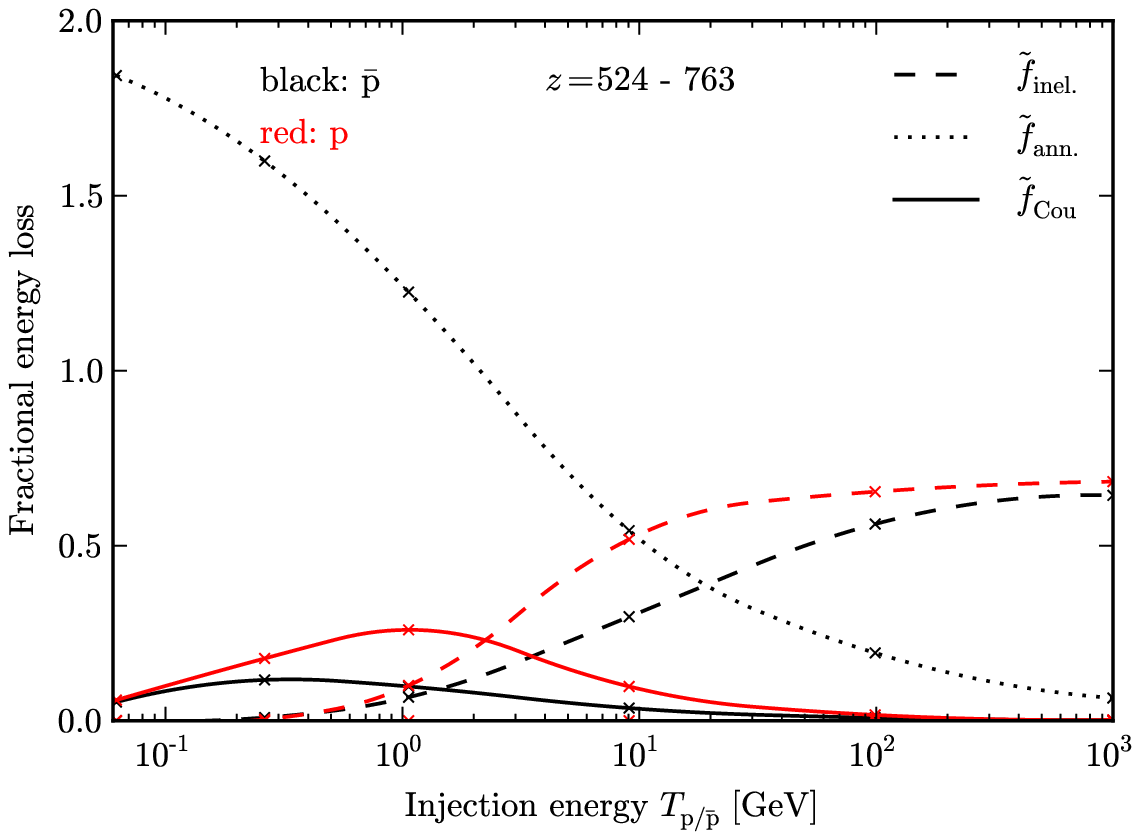}
    \includegraphics[width=\columnwidth]{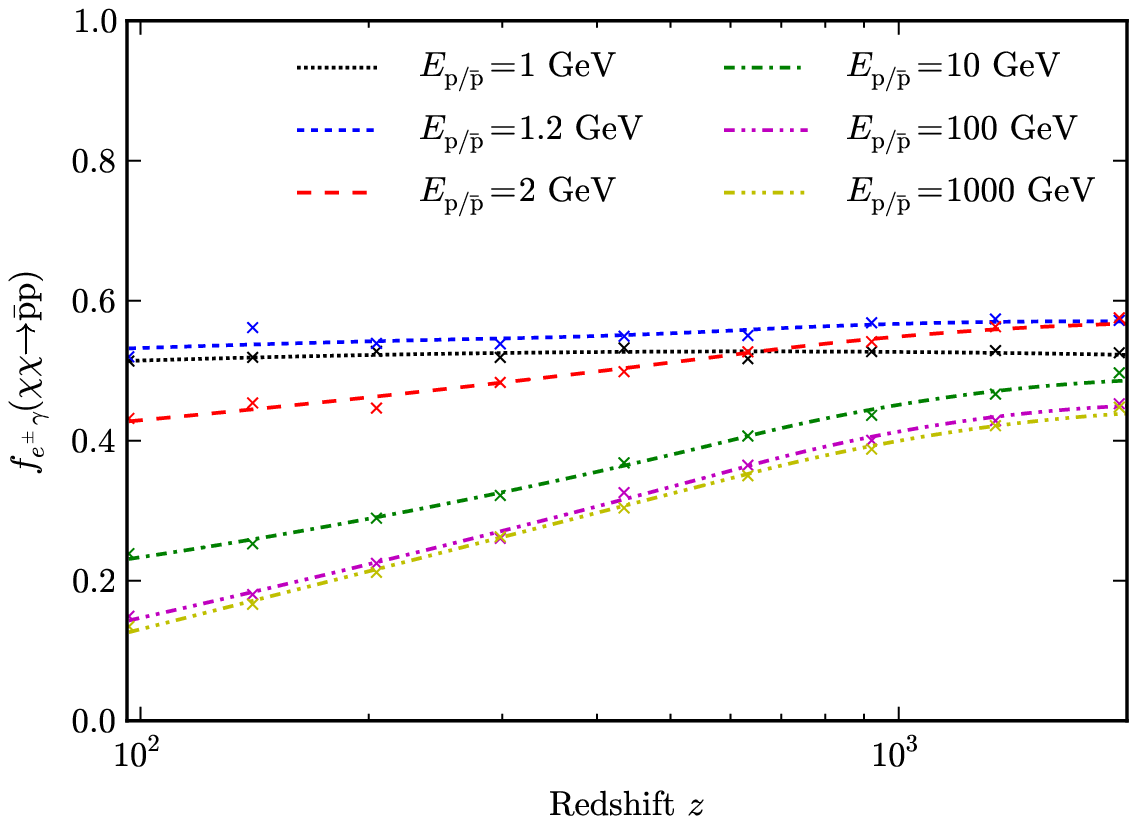}
  \end{center}
  \caption{\emph{Upper panel:} Fractional energy losses at redshift $z\sim
  600$, for different kinetic injection energies $T_{\rm p/\bar p}$. We show
  the contributions from annihilating and non-annihilating hadronic
  scatterings and from Coulomb losses separately. \emph{Lower panel:}
  Electromagnetic (i.e.~energetic $e^\pm$ and photons) energy released during
  \ppbar~propagation, as function of redshift $z$, for the toy process
  $\chi\chi\to\rm\bar p p$ with different dark matter masses $m_\chi=1.0\rm\
  GeV$ to $m_\chi=1\rm\ TeV$. In all cases we show Monte Carlo results
  (crosses) as well as interpolations with splines of degree two (lines).}
  \label{fig:results}
\end{figure}

In Fig.~\ref{fig:results} we show  the main results of this paper. In the top
panel, the resulting different components of the \emph{fractional energy loss}
of (anti-)protons are shown as function of their injection energy at redshift
$z\sim600$. Notably, the annihilation of anti-protons releases also the energy
of the target proton, which yields values $\tilde f_{\rm ann.}>1$ when
considering the anti-proton channel alone. The bottom panel shows the fraction
of the injection energy that is channeld into electromagnetic energy,
$f_{e^\pm\gamma}$, as function of redshift $z$ and for different kinetic
injection energies $T_{\rm p/\bar p}$.  For energies $T_{\rm p/\bar p}<10\ \rm
GeV$ ($T_{\rm p/\bar p}>10\ \rm GeV$) about $f_{e^\pm\gamma}\sim50\%$ (40\%)
of the energy fraction channelled into nucleons will eventually by released as
electromagnetic energy at the most relevant redshifts $z=500$--$700$.

Most importantly, for DM masses below $\sim100$ GeV, the
$f_{e^\pm\gamma}(z)$'s are approximately \emph{independent of redshift}. This
implies that the released electromagnetic energy is simply a constant
correction to the prompt electromagnetic energy directly released during DM
annihilation.  In the current literature (see
e.g.~Ref.~\cite{Slatyer:2009yq}), the fraction of the total injected energy
finally deposited in the gas as heat, ionization and excitation energy is
denoted by $f(z)$, and determined neglecting annihilation channels into
hadronic final states.  As discussed above, CMB bounds are to a good
approximation calorimetric, and the precise form of the spectrum of
electromagnetic particles can be neglected at first order in determining the
ultimate fate of the energy injected.\footnote{At low energies, the precise
assumptions of the recombination model \emph{do} become relevant and require
a proper treatment (see e.g.~CosmoRec~\cite{CosmoRec}).
A thorough study of this issue has been recently presented in Ref.~\cite{Galli:2013}.} 
We can then give a simple recipe to correct these $f$'s:
\begin{equation}
  f(z)\to f(z)\left(1+f_{e^\pm \gamma} \frac{E_{\rm N\bar N
  }}{E_{\gamma+e^\pm}}\right),\label{summeq}
\end{equation}
where $E_{\rm N\bar N}$ and $E_{\gamma+e^\pm}$ denote the prompt energy in
nucleons and in electromagnetic energy, respectively, produced during an
annihilation.  As shown in the bottom panel of Fig.~\ref{fig:results}, we find
that $f_{e^\pm\gamma}\simeq0.5$ for injected \ppbar~energies well below
$\sim10$ GeV, i.e.~for most DM masses of interest.
\bigskip
 
Throughout, we neglected \emph{elastic} \ppbar~scattering, which is well
justified for our purposes: At energies below $T_{\rm p}<1\rm \ GeV$, elastic
p-p scattering will only redistribute energy that finally is lost via Coulomb
scattering. At energies above $T_{\rm p}>1\rm \ GeV$, elastic p-p scattering
is subdominant ($\lesssim50\%$ of the total p-p cross-section); even rare
collisions with maximal energy transfer do not alter the deposition history
drastically since the energy dependence shown in Fig.~\ref{fig:results} is
weak. In case of $\rm\bar p$-p scattering, the elastic cross section is
smaller than $40\%$ at all energies; the main effect of collisions with large
energy transfer would be a slight increase in anti-proton annihilation rate.

\section{Conclusions}\label{conclusions}
It has been recently recognized that non-standard sources of heating and
ionization of the IGM can be probed via CMB anisotropies. This effect has been
used to derive bounds on DM annihilation cross-section and decay lifetime. To
a very good approximation, these bounds do not depend on ``astrophysics''
(structure formation, DM halo structure, star formation etc.).  The nature of
this probe, whose robustness  relies essentially on the accuracy of the
description of atomic and nuclear/particle physics, motivates efforts to
assess and improve the error budget of bounds  obtainable with this technique.

In astroparticle physics,  energy losses of high energy nucleons propagating
in the intergalactic medium are usually neglected outside the realm of
ultra-high energy cosmic rays. Energetic nucleons propagating in a high-$z$
cosmological medium over Hubble times, on the other hand, do suffer
significant energy losses. A physically interesting example is provided by
nucleon by-products of dark matter annihilation, which  can release a
significant fraction of their energy at high-$z$ ($\simeq {\cal O} (10^3)$).

The aim of this paper was to provide a first estimate of this process, which
goes in the direction of improving the CMB bounds for a number of channels.
Under reasonable approximations, we found that the Eq.~(\ref{summeq}) is a
fair description of the main effect, for a value $f_{e^\pm\gamma}\simeq 0.5$
over a quite large parameter space. We expect thus that for DM annihilation
channels into gluons, gauge bosons and quarks for which $ \frac{E_{\rm N\bar N
}}{E_{\gamma+e^\pm}}\simeq 0.2$, this should change the CMB bounds on DM at
the 10\% level and should be included in forthcoming Planck data analyses.

\acknowledgments
We would like to thank M. Cirelli useful comments.  At LAPTh, this activity
was developed coherently with the research axes supported by the labex grant
ENIGMASS. Partial support of the ANR grant DMAstroLHC is also acknowledged by
PS. GB acknowledges the support of the European Research Council through the
ERC Starting Grant "WIMPs Kairos".


\begin{thebibliography}{00}

\bibitem{reviews}
  G.~Jungman, M.~Kamionkowski and K.~Griest,
  Phys.\ Rept.\  {\bf 267} (1996) 195
  [hep-ph/9506380];
  L.~Bergstr\"om,
  Rept.\ Prog.\ Phys.\  {\bf 63} (2000) 793
  [hep-ph/0002126];
  C.~Munoz,
  Int.\ J.\ Mod.\ Phys.\ A {\bf 19} (2004) 3093
  [hep-ph/0309346]; 
  G.~Bertone, D.~Hooper and J.~Silk,
  Phys.\ Rept.\  {\bf 405} (2005) 279
  [hep-ph/0404175]; 
G. Bertone ({\it ed.}), ``Particle Dark Matter: Observations, Models and Searches", Cambridge University Press (2010), ISBN:9780521763684 


\bibitem{Chen:2003gz} 
  X.~-L.~Chen and M.~Kamionkowski,
  Phys.\ Rev.\ D {\bf 70}, 043502 (2004)
  [astro-ph/0310473].
  
\bibitem{Padmanabhan:2005es} 
  N.~Padmanabhan and D.~P.~Finkbeiner,
  Phys.\ Rev.\ D {\bf 72}, 023508 (2005)
  [astro-ph/0503486].
  
\bibitem{Kanzaki:2008qb} 
  T.~Kanzaki and M.~Kawasaki,
  Phys.\ Rev.\ D {\bf 78}, 103004 (2008)
  [arXiv:0805.3969 [astro-ph]].
  
\bibitem{Galli:2009zc} 
  S.~Galli, F.~Iocco, G.~Bertone and A.~Melchiorri,
  Phys.\ Rev.\ D {\bf 80}, 023505 (2009)
  [arXiv:0905.0003 [astro-ph.CO]].

\bibitem{Slatyer:2009yq} 
  T.~R.~Slatyer, N.~Padmanabhan and D.~P.~Finkbeiner,
  Phys.\ Rev.\ D {\bf 80}, 043526 (2009)
  [arXiv:0906.1197 [astro-ph.CO]].

\bibitem{Huetsi:2009ex}
  G.~H\"utsi, A.~Hektor and M.~Raidal,
  Astron.\ Astrophys.\  {\bf 505} (2009) 999
  [arXiv:0906.4550 [astro-ph.CO]].

\bibitem{Cirelli:2009bb} 
  M.~Cirelli, F.~Iocco and P.~Panci,
  JCAP {\bf 0910}, 009 (2009)
  [arXiv:0907.0719 [astro-ph.CO]].
  
\bibitem{Kanzaki:2009hf} 
  T.~Kanzaki, M.~Kawasaki and K.~Nakayama,
  Prog.\ Theor.\ Phys.\  {\bf 123}, 853 (2010)
  [arXiv:0907.3985 [astro-ph.CO]].

\bibitem{Hutsi:2011vx}
  G.~H\"utsi, J.~Chluba, A.~Hektor and M.~Raidal,
  Astron.\ Astrophys.\  {\bf 535} (2011) A26
  [arXiv:1103.2766 [astro-ph.CO]].
  
\bibitem{Galli:2011rz} 
  S.~Galli, F.~Iocco, G.~Bertone and A.~Melchiorri,
  Phys.\ Rev.\ D {\bf 84}, 027302 (2011)
  [arXiv:1106.1528 [astro-ph.CO]].

\bibitem{Finkbeiner:2011dx} 
  D.~P.~Finkbeiner, S.~Galli, T.~Lin and T.~R.~Slatyer,
  Phys.\ Rev.\ D {\bf 85}, 043522 (2012)
  [arXiv:1109.6322 [astro-ph.CO]].
  
\bibitem{Giesen:2012rp} 
  G.~Giesen, J.~Lesgourgues, B.~Audren and Y.~Ali-Haimoud,
  JCAP {\bf 1212}, 008 (2012)
  [arXiv:1209.0247 [astro-ph.CO]].
  
\bibitem{Evoli:2012qh} 
  C.~Evoli, S.~Pandolfi and A.~Ferrara,
  arXiv:1210.6845 [astro-ph.CO].

\bibitem{Slatyer:2012yq}
  T.~R.~Slatyer,
  arXiv:1211.0283 [astro-ph.CO].

\bibitem{Cline:2013fm} 
  J.~M.~Cline and P.~Scott,
  JCAP {\bf 1303}, 044 (2013)
  [Erratum-ibid.\  {\bf 1305}, E01 (2013)]
  [arXiv:1301.5908 [astro-ph.CO]].

\bibitem{Chluba:2009uv}
  J.~Chluba,
  arXiv:0910.3663 [astro-ph.CO].
  
  
\bibitem{Chluba:2010ca} 
  J.~Chluba and R.~M.~Thomas,
Mon.\ Not.\ Roy.\ Astron.\ Soc.\  {\bf 412}, 748 (2011).
    [arXiv:1010.3631 [astro-ph.CO]].
  
\bibitem{Cirelli:2010xx} 
  M.~Cirelli, G.~Corcella, A.~Hektor, G.~H\"utsi, M.~Kadastik, P.~Panci, M.~Raidal and F.~Sala {\it et al.},
  JCAP {\bf 1103}, 051 (2011)
  [Erratum-ibid.\  {\bf 1210}, E01 (2012)]
  [arXiv:1012.4515 [hep-ph]].

\bibitem{Jackson:1998}
  J.~D.~Jackson,
  \textit{Classical Electrodynamics Third Edition},
  John Wiley and Sons Inc (1999)

\bibitem{Jedamzik:2006xz} 
  K.~Jedamzik,
  Phys.\ Rev.\ D {\bf 74}, 103509 (2006)
  [hep-ph/0604251].
  
\bibitem{Moskalenko:2001ya} 
  I.~V.~Moskalenko, A.~W.~Strong, J.~F.~Ormes and M.~S.~Potgieter,
  Astrophys.\ J.\  {\bf 565}, 280 (2002)
  [astro-ph/0106567].
  
  \bibitem{Evoli:2012zz} 
  C.~Evoli, M.~Valdes, A.~Ferrara and N.~Yoshida,
  Mon.\ Not.\ Roy.\ Astron.\ Soc.\  {\bf 422}, 420 (2012).

\bibitem{Steigman:1976ev} 
  G.~Steigman,
  Ann.\ Rev.\ Astron.\ Astrophys.\  {\bf 14}, 339 (1976).

\bibitem{CosmoRec}
\url{http://www.cita.utoronto.ca/~jchluba/Science_Jens/Recombination/CosmoRec.html}.

\bibitem{Galli:2013} 
  S.~Galli, T.~R.~Slatyer, M.~Valdes and F.~Iocco,
  arXiv:1306.0563 [astro-ph.CO].

\end{thebibliography}
\end{document}